\documentclass[12pt]{iopart}
\usepackage{graphicx}
\usepackage{amsfonts,amssymb,bm,ulem}

\begin{document}
\title{Low temperature transport properties of multigraphene structures
 on 6H-SiC obtained by thermal graphitization: evidences of a presence of nearly perfect graphene
 layer.}
\author{A. A.~Lebedev,$^1$ N. V.~Agrinskaya,$^1$ V. A.~Beresovets,$^{1,2}$
V. I.~Kozub,$^1$ S. P.~Lebedev,$^1$,A.A.~Sitnikova,$^1$}

\address{$^1$A. F. Ioffe Institute, Russian Academy of Sciences,
  194021 Saint Petersburg, Russia;
$^2$International Laboratory of High Magnetic Fields and Low
Temperatures, 95 Gajowicka str. 53421 Wroclaw, Poland }

\begin{abstract}

Transport properties of multigraphene layers on 6H-SiC substrates
fabricated by thermal graphitization of SiC were studied. The
principal result is that these structures were shown to  contain a
nearly perfect graphene layer situated between the SiC substrate
and multgraphene layer.  It was found that the curves of
magnetoresistance and Shubnikov- de Haas oscillations shown the
features, typical for single-layered graphene. The low temperature
resistance demonstrated an increase with temperature increase,
which also corresponds to a behavior typical for single-layered
graphene (antilocalization). However at higher temperatures the
resistance decreased with an increase of temperature, which
corresponds to a weak localization. We believe that the observed
behavior can be explained by a parallel combination of
contributions to the conductivity of single-layered graphene and
of multigraphene, the latter allowing to escape damages of the
graphene by atmosphere effect.

\end{abstract}

\section{
Introduction}

At last time one of the popular methods of graphene layers
formation is thermal graphitization of monocrystalline silicon
carbide at vacuum or in argon atmosphere \cite{ber}. This method
gives a possibility - in contrast to standard approaches - to
fabricate large graphene samples, allowing the following
lithography treatment. The essence of this method is
non-stoichiometric vaporing of silicon from a surface of
mono-crystalline SiC at high temperature heating leading to a
formation of hexagonal lattice of the rest carbon atoms at its
surface. A quality of the layers obtained is controlled by methods
of pre-grow treatment of the SiC substrate. Note that, as we know,
this method most probably leads to a formation of multilayered
structure. Thus we believe that the study of such structures is an
important problem for the graphene physics and technology.

In our previous paper \cite{ours1} we studied structure and
transport properties of multigraphene layers, subjected to
pre-grow thermal treatment. The obtained multigraphene layers had
properties of 2D hole gas with relatively high carrier
concentration ($10^{12} cm^{-2}$) and low mobility 100 $cm^2/( V
s)$. Temperature curves of resistance and magnetoresistance
(including low field negative magnetoresistance peak) demonstrated
weak localization effects. However we have not observed specific
transport properties related to a nature of carriers in
single-layered graphene and to a specific density of states
spectrum \cite{2} , like weak antilocalization (WAL) and
Shubnikov- de Haas oscillations.

In this paper we studied transport properties of more perfect
graphene layers on SiC substrate, grown by sublimation in vacuum.
In this case the transport properties of the obtained samples
clearly demonstrated the features, typical for graphene layers. In
particular, it was shown, that the behavior observed can be
explained only as a result of combination of parallel
contributions to conductivity of single-layered graphene and of
multigraphene. This conclusion seems to be of principal importance
since the structure under study is shown to contain nearly perfect
graphene layer isolated from atmosphere by the multigraphene. The
latter has much less conductivity with respect to the graphene
layer and thus does not affect significantly the properties of the
graphene.

\section{\bf Experiment}

For studies of the transport properties of graphene films we used
samples obtained on the surface of monocrystalline silicon carbide
by sublimation in vacuum \cite{leb1}. As the substrate we used
semiinsulating plates 6H-SiC fabricated by CREE Inc. The graphene
film was grown on the face $C_{(000\bar{1})}$. Directly before a
growth of the graphene film the standard procedure of the washing
of substrates in organic solvents was applied. To remove the
surface layer disturbed by polishing of the plates we applied a
technology of pre-growth treatment of SiC substrates developed
earlier \cite{leb2}. This technology allows to improve
significantly a quality of the grown graphene with respect to the
growth on an untreated substrate. A growth of graphene films was
carried out in high-vacuum chamber with a residual  pressure
$~10^{-6}$ atm.. The growth temperature was 1400-1500 C, the
growth time was 15-30 min. A quality of the grown films was
controlled "ex-situ" with a help of atomic-force microscopy (AFM)
and of Raman spectroscopy. After the growth structures of the Hall
bar geometry were fromed on the graphene film (a distance between
the probes $L = 20 \mu m$, a width of the graphene strip was W =
$10\mu m$).

This structure was investigated with a help of a transparent
high-resolution electron microscopy. The sample for investigation
was prepared according to a standard method including a mechanic
polishing of the transversal cut with a following ion etching. The
investigations were made with a help of a transparent electron
microscope. As a result, a typical image of a transversal cut of
the structure graphene-SiC was obtained (see Fig.1).

\begin{figure}[htbp]
    \centering
        \includegraphics[width=0.45\textwidth]{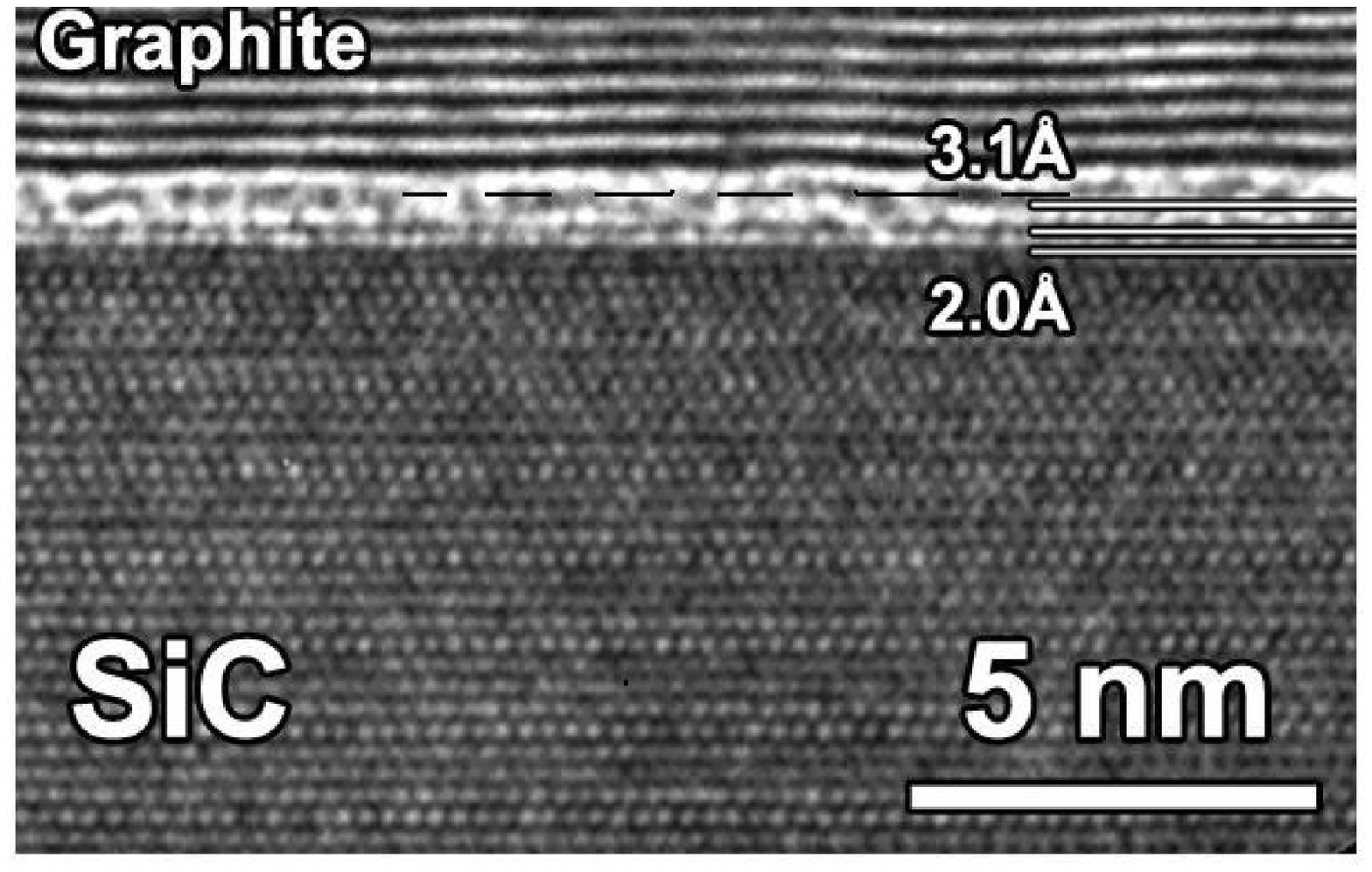}
        \caption{\ref{fig1}.

The image of a transversal cut of the structure graphene/SiC,
obtained with a help of a high resolution transparent electron
microscopy. A dotted line represents the graphene layer.
        1.}
    \label{fig:fig1}
\end{figure}

The first graphene layer (layer 0) was situated directly on the
surface of silicon carbide. Note that the distance between the
last layer of the silicon carbide and the first graphene layer was
only 2A. This fact evidences a presence of a covalent binding
between Si atoms of the last layer of SiC and C atoms of the first
graphene layer. As it was shown in \cite{TEM}, the first layer has
a function of the buffer layer and has a semiconducting character
(giving a small contribution to conductivity at low temperatures).
The next graphene layer (layer 1) is situated at a distance about
3.1 A from the layer 0. The next layers (including 8 layers) are
situated at distances $3,39$ A from each other. This value is
close to the value of the interlayer distance in graphite and is
characteristic for the Van der Vaalse bondings. Thus, the
electron-microscopic studies imply a presence of 3-layered
structure: layer 0 - the buffer layer, not participating in the
conductance, layer 1 - the layer of "perfect" graphene and the
multigraphene layer (including 8 layers).

The galvanomagnetic effects studies were made on the sample with
standard Hall geometry and included three orientations of the
sample with respect to the magnetic field $\bf B$ (${\bf n}
\parallel {\bf B} \perp {\bf j}$, ${\bf n }\perp  {\bf B} $ $
\perp {\bf j}$ è ${\bf n} \perp {\bf B}
\parallel {\bf j}$),where $\bf n$ and $\bf j$ are the vectors
of the normals with respect to the layers planes and to the
current density, respectively. At magnetic fields 0-15 T we
measured the Hall effect and magnetoresistance. According to the
sign of the Hall effect we had specified the hole type of the
conductivity at the whole region of temperatures (1,5 - 130 Ê).
The Hall hole concentration, extracted at weak magnetic fields (up
to B = 1 T) appeared to be equal to $ð= 8,3\cdot 10^{12} cm^{-2}$
while at high magnetic fields ($B$ up to 15 T) $ð=1,1 \cdot
10^{13} cm^{-2}$ at T=4,2Ê. The calculated value of the Hall
mobility appeared to be  $77 cm^2V^{-1}s^{-1}$ at $T$ = 1.5 K,
which is much smaller than in single-layered epitaxial graphene
(5000-10000 $cm^2V^{-1}s^{-1}$)
 \cite{2} .

The general pattern of the magnetoresistance (MR)  curves at the
fields up to 5 T is present on Fig.1

\begin{figure}[htbp]
    \centering
        \includegraphics[width=0.45\textwidth]{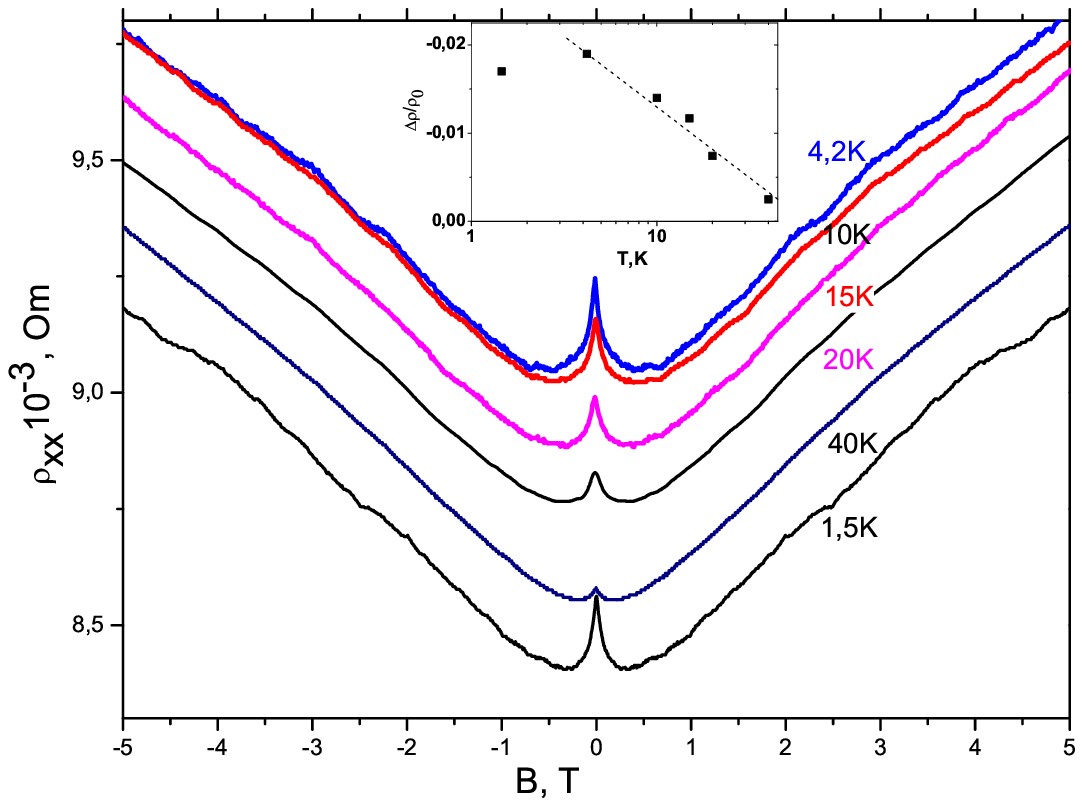}
        \caption{ \ref{fig:fig2}.

The curves of magnetoresistance at different temperatures. At the
insert - the temperature dependence of the negative MR peak.

        }
    \label{fig:fig2}
\end{figure}

It is seen, that at high fields ($B > 1$ T)  at low temperatures
($T < 15$ K), a non-monotonous contribution to the signal had
appeared. The corresponding oscillations were periodic on the
scale of the inverse fields (SdH oscillations). In the case of the
sample plane orientations parallel to the the axis of the magnetic
field (${\bf n} \perp {\bf B} \perp {\bf j}$ è ${\bf n} \perp {\bf
B} \parallel {\bf j}$) the SdH oscillations were absent which
evidences the 2D character of conductivity of the studied samples.

At the region of weak magnetic fields ($B < 0.5$ T) the effect of
negative magnetoresistance (NMR) was found, Fig. 1. The
temperature behavior of the NMR peak (insert in Fig.1) appeared to
be proportional to $ln T$, and thus it can pe ascribed to the weak
localization effect (WL) for the diffusive transport of 2D
carriers. Such an interpretation of this peak is supported by the
following facts:

1) An absence of the WL efect for the magnetic field directions
parallel to the plane of the sample,

2) A decrease of the WL effect with an increase of temperature
proportionally to $ln T$ up to its complete suppression at $T > 50
$ K.

With a decrease of temperature up to 1.5 K an increase of the  WL
peak is replaced by its suppression. At magnetic fields higher
than 0.3 T the crossover to the positive magnetoresistance (PMR)
is observed. The observed  crossover with an increase of the
magnetic field can be related to a manifestation of the
anti-localization addition to the classical Drude conductivity.
Such a behavior is similar to the magnetoresistance related to the
weak antilocalization, observed in \cite{2} for epitaxial
single-layered graphene, grown on 4H-SiC.

     The temperature dependence of resistance, Fig. 3, at the
     region 10- 100 K is also described by the logarithmic law,
     typical for WL ( the curves slope gives for the
     prelogarithmic factor a resistance value about 3 kOhm (which
     is close enough to $\hbar/e^2$).
    \begin{figure}[htbp]
    \centering
        \includegraphics[width=0.47\textwidth]{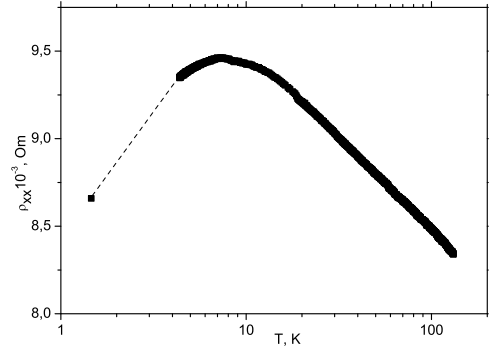}
        \caption{ \ref{fig:fig3} The temperature dependence of resistance
        .
         }
\label{fig:fig3}
\end{figure}
At Fig. 2 it is seen, that at $T < 7$ K a reversal of the sign of
the slope of temperature behavior (to the metallic behavior) is
observed. This could be related to the manifestation of the WAL in
systems with spin-orbital coupling \cite{3}. However, acording to
the theoretical considerations concerning the nature of carriers
in single-layered graphene and its specific density of state
spectrum \cite{2}, in graphene WAL has not the spin-orbital
character, but is rather controlled by the intervalley scattering.
Experimentally, WAL had been observed in saingle layered graphene
magnetoresistance at stronger magnetic fields, than WL \cite{4}.
In our sample the MR curves (Fig.1) also demonstrate a transition
to the PMR at the fields $0,2T>B>0,5T$, which can be also related
to a manifestation of WAL. However, as we know, until now a
transition to the metallic behavior of conductivity in
single-layered graphene at low temperatures and $B = 0$ was not
observed. An exclusion is the paper, where such a quasimetallic
behavior was observed at thick enough multigraphene layers (20 nm)
and was ascribed to a contribution of the interfaces \cite{5}.

At high fields the SdH oscillations were observed
 \begin{figure}[htbp]
    \centering
        \includegraphics[width=0.47\textwidth]{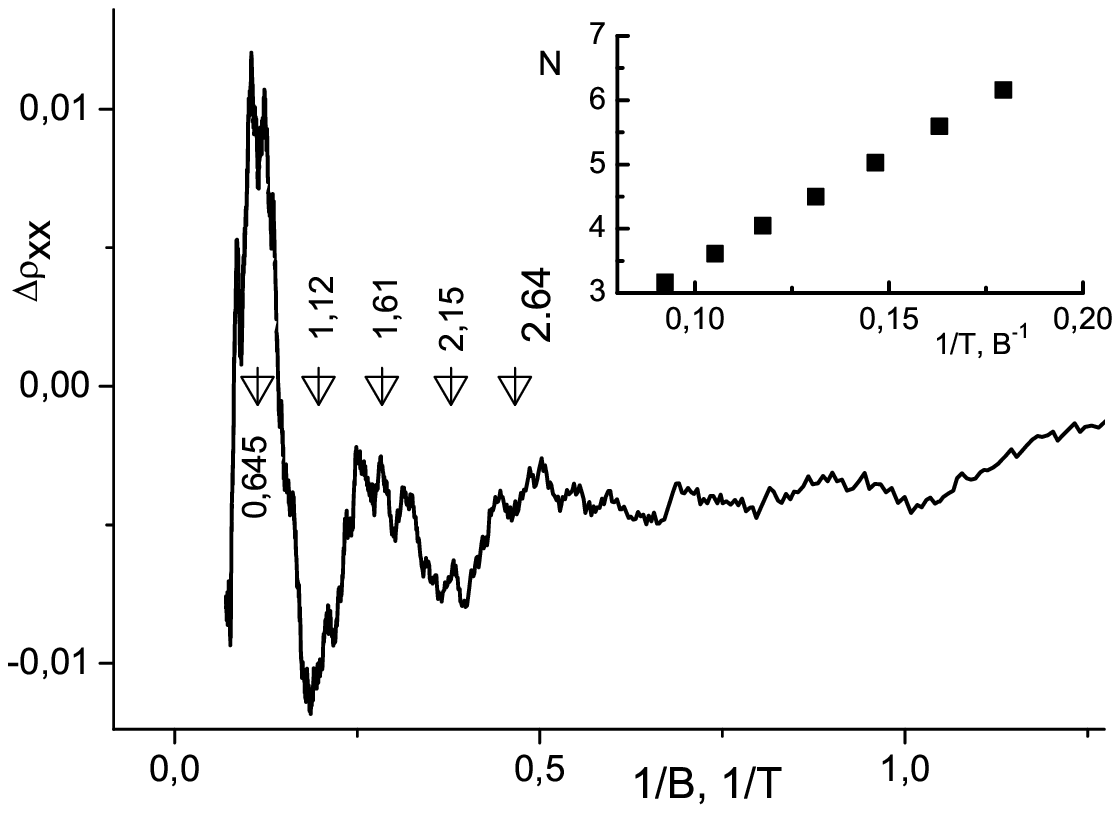}
        \caption{
        \ref{fig:fig4}.

A curve of magnetoresistance in the scale of inverse magnetic
fields, obtained by a substraction of the background MR (described
by a polinomial of a second order). The insert gives the Landau
numbers as a function of the positions of SdH maxima,
corresponding to the oscillations of the second type, observed at
high magnetic fields).}

\label{fig:fig4}
\end{figure}

The Fig. 3 presents a resistance behavior as a function of
magnetic field, obtained by a substraction from the MR curve of
the polinomial of the second order. It is seen, that in terms of
the inverse magnetic fields the pronounced oscillations with a
period $\triangle=0,1762 1/T$ are observed. The value of the
carrier concentration, extracted from the period of the
low-temperature SdH oscillations with an account of spin and value
splitting for the parabolic spectrum gives $p=
 2e/\triangle\hbar\pi= 2,7\cdot 10^{11} cm^{-2}$.
In addition to these oscillations, the Fig.3 also demonstrates
weaker structure, which is pronounced only up to temperatures 30
K. When separating these oscillations, one also find that these
are periodic in terms of the inverse magnetic field, and its
period is $\triangle=0,029 1/T$) (see insert in Fig. 3). This
corresponds to the carrier concentration $p= 1,7\cdot 10^{12}
cm^{-2}$ which is by an order of magnitude higher than the
concentration mentioned above. While these oscillations are
observed at higher magnetic fields, one can expect that they are
controlled by a layer with a lower conductivity (multigraphene).

\section{Discussion}

The experimental results presented above evidence, on the one
hand, that the system under study manifests properties, typical
for "dirty" metal (weak localization and relatively small
mobilities). On the other hand, we observed weak antilocalization
and SdH oscillations, which is typical for clean single-layered
graphene. These facts allows to suggest, that in the system, in
addition to the layer of perfect graphene, also exists a layer of
multigraphene with large number of defects. We believe that the
most unexpected feature is the "reversed" temperature dependence
of resistance with respect to the one common for graphene. As it
is known, in the first approximation the temperature dependent
contribution to the graphene resistance has a form
\begin{equation}\label{1}
2(1/\pi h)\ln (\tau_{\varphi}/\tau_{tr})
\end{equation}
where $\tau_{\varphi}(T)$ is the phase breaking time, $\tau_{tr}$
is the transport relaxation time in graphene while factor 2 is
related to independent contributions of the two valleys. Here it
is assumed that $\tau_{\varphi} < \tau_{iv1} $, where $\tau_{iv1}$
is the intervalley scattering time in the graphene layer which is
expected to be large enough. Indeed, this time is related to the
acts of scattering involving large change of the wave vector (of
the order of the vector of the reciprocal lattice).
Correspondingly, it is expected that at reasonable temperatures
the contribution in question increases with a decrease of
temperature (antilocalization). Note however that the temperature
dependence of $\tau_{tr}$ can lead to violation of this
conclusion. Namely, while during the temperature decrease one
expects a saturation of $\tau_{tr}$, the ratio
$\tau_{\varphi}/\tau_{tr}$  can, in principle, decay with the
temperature increase, which leads to non-monotonous temperature
behavior of the logarithm argument. However, this factor still
does not allow to explain other features of the system in
question. Thus let us assume, that in addition to the layer of
"perfect" graphene our system also contains a conductive layer
with small values of $\tau_{iv}$ - in particular, formed from
multilayer material with different position of the cells and with
klarge number of interlayer defects. In this case at high enough
temperatures the situation $\tau_{iv,2} < \tau_{\varphi,2}$ can be
realized. Here we denoted parameters corresponding to the second
layer by index 2 while in the following we denote the parameters
of the graphene layer by index 1. The contribution of this second
layer to conductivity is controlled by the weak localization,
since there is a strong mixing between the valleys, and the
corresponding contribution can be described as
\begin{equation}\label{2}
-   (1/\pi h)\ln (\tau_{\varphi_2}/\tau_{tr,2})
\end{equation}
Thus we simultaneously have two contributions, which can manifest
itself at different temperatures: the contribution of weak
localization, related to the defect layer, and the contribution of
antilocalization, related to more perfect graphene layer. Having
in mind a summation of the two contributions, we obtain for the
interference addition
\begin{equation}\label{3}
(1/\pi h) \ln \left(
(\tau_{\varphi1}/\tau_{tr1})^2(\tau_{tr2}/\tau_{\varphi2})\right)
\end{equation}
As it is seen, at low temperatures the dominant contribution of
the antilocalization is expected, while at higher temperatures -
the dominant contribution of the weak localization. To estimate
the temperature behavior of conductivity, let us consider the
situation in more detail, calculating the temperature derivative
of the interference addition:
\begin{equation}\label{4}
2\frac{\tau_{tr1}}{\tau_{\varphi1}}\frac{\partial(\tau_{\varphi1}/\tau_{tr1})}{\partial
T} -
\frac{\tau_{tr2}}{\tau_{\varphi2}}\frac{\partial(\tau_{\varphi2}/\tau_{tr2})}{\partial
T}
\end{equation}
According to the model suggested ("pure" graphene layer shunted by
"dirty" multigraphene layer) one can make the following
assumptions. First, one expects that $\tau_{tr1} > \tau_{tr2}$.
Secondly, the behavior of $\tau_{\varphi2}$ is typical for
disordered materials and exhibits a saturation with temperature
decrease (see, e.g.,  \cite{Galperin},\cite{KozubAleshin}). Thus
one can expect that at low temperatures the temperature derivative
is controlled by the first term (antilocalization related to the
graphene layer) while the second term is suppressed by the weak
dependence of $\tau_{\varphi2}(T)$. However with temperature
increase the dominant role is played by the second term (weak
localization related to the multigraphene layer). Such a behavior
is emphasized by the relative small value of $\tau_{tr2}$. It is
clear that the conductivity minimum (and thus resistivity maximum)
is reached when the expression of \ref{4} vanishes. One can also
expect, that the approximate estimate of the minimum temperature
corresponds to approximate equality between the contributions of
the localization and of the antilocalization.

Note that the further decrease of temperature, leading to
violation of the unequality $\tau_{\varphi1} < \tau_{iv1}$ can
lead to an increase of resistance with temperature decrease, that
is to restoration of the weak localization.

As for magnetoresistance, it is natural to expect, that at high
temperatures the contribution of weak localization is emphasized,
while at lower temperatures this contribution decreases. It is
this behavior which is demonstrated by temperature behavior of
magnetoresistance, where the low field peak of negative
magnetoresistance is suppressed with temperature decrease.

As for the observation of Shubnikov - de Haas oscillations, it
does not contradict to the model suggested. Namely, it is the
highly conducting graphene layer with carrier concentration $p=
2,7\cdot 10^{11} cm^{-2}$ which is responsible for these
oscillation, although their amplitude is partly suppressed by a
presence of the layer with low conductivity. At the same time, the
carrier concentration extracted from the Hall effect $p= 8,3\cdot
10^{12} cm^{-2}$ is formed mostly by the multigraphene layer.

Concluding, it is instructive to emphasize a role of the
multigraphene layer. On the one hand, it makes a picture of the
conductivity for the structure graphene/SiC to be more complex.
But, on the other hand, the multigraphene  play a protecting role
for the graphene layer, preventing its contact with the
atmosphere. As it is known, the effect of atmosphere, leading, in
particular, to a formation of background dopants, gives a
significant limitation to application of single-layered graphene.
Then, the multigraphene provides a good adhesy of the metal to the
graphene layer in course of fabrication of contacts. The principal
result of our studies is that for structures, produced by thermal
graphitization of SiC still contain neraly perfect graphene layer
in between of SiC substrate and multigraphene layer. of Thus, the
corresponding multilayered graphene structure can have advantages
with respect to single-layered graphene.

\section{Acknowledgements}

This work was supported by RFBR foundation (project 10-02-00544à
and project 12-02-00165à).

\end{document}